\documentclass[%
]{emulateapj}

\usepackage{graphicx}
\usepackage{dcolumn}
\usepackage{bm}
\usepackage{enumitem}
\usepackage{textcomp}

\usepackage{color}

\newcommand{\Mc}{\mathcal{M}_c}
\newcommand{\Msun}{{\rm M}_\odot}
\newcommand{\LALInf}{\emph{LALInference}}

\begin{document}


\title{Neutron stars versus black holes: probing the mass gap with LIGO/Virgo}

\author{Tyson B. Littenberg}
\affiliation{Center for Interdisciplinary Exploration and Research in Astrophysics (CIERA)
and
Department of Physics and Astronomy
Northwestern University
2145 Sheridan Road
Evanston, IL 60208
USA}
\author{Ben Farr}
\affiliation{Enrico Fermi Institute, University of Chicago, Chicago, IL 60637, USA}
\author{Scott Coughlin}
\affiliation{Cardiff University, Cardiff, CF24 3AA, United Kingdom.}
\author{Vicky Kalogera}
\affiliation{Center for Interdisciplinary Exploration and Research in Astrophysics (CIERA)
and
Department of Physics and Astronomy
Northwestern University
2145 Sheridan Road
Evanston, IL 60208
USA}
\author{Daniel E. Holz}
\affiliation{Enrico Fermi Institute, Department of Physics, and Kavli Institute for Cosmological Physics, University of Chicago, Chicago, IL 60637, USA}
\date{\today}

\begin{abstract}
Inspirals and mergers of black hole (BHs) and/or neutron star (NSs) binaries are expected to be abundant sources for ground-based gravitational-wave (GW) detectors.
We assess the capabilities of Advanced LIGO and Virgo to measure component masses using inspiral waveform models including spin-precession effects using a large ensemble of GW sources {\bf randomly oriented and distributed uniformly in volume.  For 1000 sources this yields signal-to-noise ratios between 7 and 200}.  
We make quantitative predictions for how well LIGO and Virgo will distinguish between BHs and NSs and appraise the prospect of using LIGO/Virgo observations to definitively confirm, or reject, the existence of a putative ``mass gap'' between NSs ($m\leq3\ M_\odot$) and BHs ($m\geq 5\ M_\odot$).
We find sources with the smaller mass component satisfying $m_2 \lesssim1.5\ M_\odot$ to be unambiguously identified as containing at least one NS, while systems with $m_2\gtrsim6\ M_\odot$ will be confirmed binary BHs.  
Binary BHs with $m_2<5\ M_\odot$  (i.e., in the gap) cannot generically be distinguished from NSBH binaries.
High-mass NSs ($2<m<3$ $M_\odot$) are often consistent with low-mass BH ($m<5\ M_\odot$), posing a challenge for determining the maximum NS  mass from LIGO/Virgo observations alone.
Individual sources will seldom be measured well enough to confirm objects in the mass gap and statistical inferences drawn from the detected population will be strongly dependent on the underlying distribution.
If nature happens to provide a mass distribution with the populations relatively cleanly separated in chirp mass space, as some population synthesis models suggest, then NSs and BHs are more easily distinguishable.
\end{abstract}

\maketitle


\section{\label{sec:intro}Introduction}

Advanced LIGO~\citep{aLIGO} and Advanced Virgo~\citep{aVirgo} will be the most sensitive observatories in the gravitational-wave (GW) spectrum between 10 Hz to a few kHz.  
Binary systems comprised of compact stellar remnants, such as stellar mass black holes (BH) and neutron stars (NS) merge at frequencies between $\sim\,100$ and $\sim\,1000$ Hz and are the primary science target for the LIGO/Virgo (LV) network.

GW observations will investigate these systems in ways not accessible to electromagnetic observations.
Detailed observations of individual sources will provide unparalleled insight into strong field gravity and the NS equation of state~\citep{Flanagan:2007ix, Read:2009yp, Lackey:2013axa, Wade:2014vqa, Li:2011cg, Sampson:2013lpa}.
Inferred characteristics of the compact binary population will feed back into complicated, ill-constrained physics of compact object formation and binary evolution~(\citet{Abadie:2010cf} and references therein).

We investigate the capabilities of an Advanced LV network of detectors to constrain the individual component masses of a compact binary and thereby distinguish between NSs and BHs.  We identify the mass intervals for which neutron star black-hole binaries (NSBH), and binary black hole (BBH) systems can be securely distinguished.  This is the first large-scale study to characterize compact binary posterior distribution functions (PDFs) including spin-precession effects over a broad range of masses and spins.

We use our results to assess LV's role in resolving the debate over the compact object ``mass gap.''
Observations of X-ray binaries suggest a depletion in the mass distribution of compact remnants between the highest mass NSs ($\sim 2 \Msun$) and the lowest mass BHs ($\gtrsim 5 \Msun$) ~\citep{Ozel:2010su, Farr:2010tu}.  
Inferring masses electromagnetically is challenging and systematic errors may dominate.  Including variable emission from the accretion flow in the analysis of the same X-ray binaries systematically finds lower masses and disfavors the mass gap~\citep{Kreidberg:2012ud}.

The mass distribution of NSs and BHs has implications on plausible explanations for the core-collapse supernova mechanism.  
\citet{Belczynski:2011bn} suggests the mass gap as observational evidence that supernovae develop rapidly (within 100 to 200 ms), while a ``filled gap'' favors longer timescale explosion mechanisms.  
\citet{Kochanek:2013yca} and \citet{Clausen:2014wia} find that a bimodal mass distribution for compact remnants is a natural consequence of high mass red supergiants ending their lives as failed supernovae. 

GW observations have been suggested as a means of resolving this controversy because the component masses are directly encoded in the signal.
We use our study to forecast whether LV observations can confirm or falsify a mass gap between NSs and BHs.
Our results are to be contrasted with the contemporaneous paper from \citet{Mandel:2015spa}, in which they assume a specific mass distribution (generated from population synthesis) and found it is possible to distinguish BHNS and BBHs and infer a gap after tens of detections.

\section{Inferring physical parameters from gravitational-wave data}\label{sec:pe}

Inspiral waveforms are well understood from post-Newtonian (PN) expansions of the binary dynamics \citep[e.g.][]{Blanchet:2006zz}.
PN waveforms enable template-base data analysis methods where many trial waveforms are compared to the data.
The model waveforms, or templates, are parameterized by the masses, spins, location, and orientation of the binary.

GW signals from compact binaries will be in the LV sensitivity band for tens of seconds to minutes, evolving through $N_{\rm cycle}\sim10^2$--$10^4$ cycles before exiting the band or merging.  
Long duration signals place strict demands on the acceptable phase difference between the template waveforms and the signal.  
To leading order in the PN expansion, the phase evolution depends only on the ``chirp mass''~\citep[see][]{Peters:1963ux} $\Mc \equiv \left(m_1 m_2\right)^{3/5}\left(m_1+m_2\right)^{-1/5}$ where $m_1>m_2$ are the component masses of the binary.
The uncertainty in $\Mc$ scales as $( N_{\rm cycle}  )^{-1/2}$~\citep{Sathyaprakash:2009xs} and is thereby constrained to high precision while the component masses are completely degenerate.  
Higher order corrections introduce the mass ratio $q=m_2/m_1$ and couplings between the intrinsic angular momentum of the component bodies $\vec{S}_{1,2}$ and the orbital angular momenta of the system $\vec{L}$ breaking the degeneracy between $m_1$ and $m_2$, though large correlations remain \citep{Cutler:1994ys, Poisson:1995ef}. 

\citet{Hannam:2013uu} investigated how well LV can distinguish between NSs and BHs by approximating parameter confidence intervals using the match between a proposed signal and a grid of templates.  
Their study used a simplified version of the full PN waveforms parameterized by a single ``effective spin''  
and concluded that most LV detections will not provide unambiguous separation between NSs and BHs.

The effective spin approximation overstates the degeneracy between mass and spin because it ignores precession of the orbital plane and spin alignments.
 \citet{Chatziioannou:2014coa} revisited the topic of component-mass determination with a Markov chain Monte Carlo (MCMC) analysis using waveforms that included spin and orbital precession .
Their study found that spin precession reduces mass-spin correlations, concluding that BNS systems with components consistent with known NSs in binaries would not be misidentified as either low-mass black holes or ``exotic'' neutron stars which they define as having masses either below $1\ \Msun$ or above $2.5\ \Msun$, and/or with dimensionless spins $\chi>0.05.$

In this work we use the same methodology of \citet{Chatziioannou:2014coa} -- employing an MCMC with templates that include full two-spin precession effects to infer the PDF -- but use a large ensemble of NSBH and BBH signals, paying particular attention to LV's ability to identify sources which have one, or both, components in the mass gap.   

\section{\label{sec:method} Method}

Our study uses the \LALInf\ software library for recovering the  parameters of compact binary systems.  
In our work we elect to use the MCMC implementation \texttt{lalinference\_mcmc} though results do not depend on the chosen sampler. 
A complete description of the software is found in~\citet{LALInference}.

\LALInf\ is part the LSC Analysis Library (LAL) which has a wide variety of template waveforms available.
{\bf For this work we use the SpinTaylorT2 waveform implemented in LAL.  A detailed description of the waveform can be found in Appendix B of \citet{Nitz:2013mxa}}.
To simulate a population of plausible LV detections we draw 1000 binary parameter combinations from a uniform distribution in component masses with $m_1\geq m_2$, $m_i\geq 1\ \Msun$ and $m_1+m_2 \leq 30\ \Msun$. 
The maximum total mass of $30\ \Msun$ is chosen so that the merger and ring-down portion of the waveform (where the PN approximation is invalid) does not dominate the signal as observed by LV.
Dimensionless spin magnitudes $\chi$ are drawn uniformly from $[0,1]$ and the spin vector orientation is randomly distributed over a sphere with respect to $\vec{L}$.  
Notice that our choice of simulated signals includes the possibility of neutron stars with anomalously high spins $\chi>0.7$~\citep{Lo:2011}.  

\begin{figure}
   \centering
   \includegraphics[width=\linewidth]{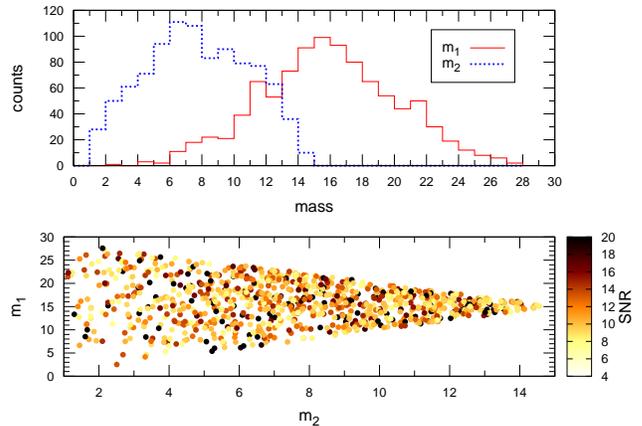} 
   \caption{\small{The mass distribution of our simulated population of compact binaries in units of $\Msun$.
   [Top panel] Histogram for $m_1$ (red, solid) and $m_2$ (blue dotted).
The bottom panel is a scatter plot of $m_1$ and $m_2$ colored by the source's SNR over the advanced LV network.}}
   \label{fig:masses}
\end{figure}

The sky location is distributed uniformly over the celestial sphere, orientations are randomly distributed, and the distance to the binary is uniform in volume.
We reject sources which do not have signal-to-noise ratio $\rm{SNR}>5$ in two or more detectors.    
Figure~\ref{fig:masses} shows the mass distributions of our population.  
The bottom panel is a scatter plot of $m_1$ and $m_2$ colored by the source SNR over the Advanced LV network.
{\bf It is important to note that our simulated population includes binaries at sufficiently high mass ($M\gtrsim12 \Msun$) that the merger and ringdown portion of the signal is detectable~\citep{Buonanno:2009zt}.
Our study is focused on quantifying how well masses can be determined purely from the inspiral part of the waveform, for which we have reliable waveforms appropriate for parameter estimation (i.e., valid across all parameter space).
Using inspiral waveforms for both our signal simulations and parameter recovery allows us to assess the inspiral effects separate from merger-ringdown effects.
However we are in urgent need of precessing waveforms that include the merger/ringdown and are valid over the full prior range to avoid systematic errors in analyses of real data.
Currently no such waveforms are available.  Available inspiral models that include precession and merger ringdown must be calibrated to numerical relativity simulations and to date are only valid at mass ratios $q>1/4$~\citep{Hannam:2013oca}.
Such limited-validity waveforms lead to non-quantified, systematic biases and cannot be used for parameter estimation at present.
In the absence of generic waveforms it may be necessary to apply a low-pass filter on real data to mask any merger and ringdown signals. }

\begin{figure}
   \centering
   \includegraphics[width=\linewidth]{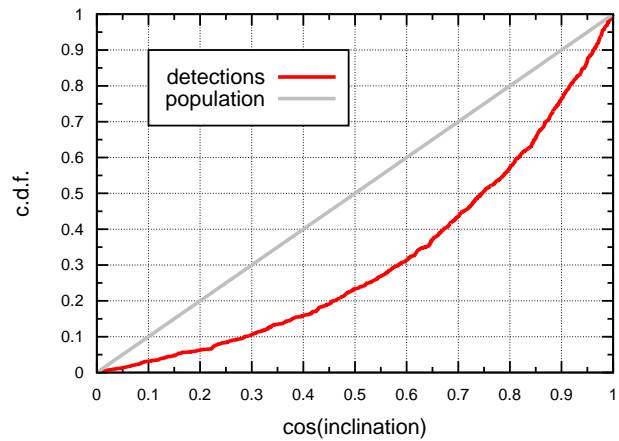} 
   \caption{\small{Cumulative distribution function for detected inclination angles (red) compared to the isotropicly oriented underlying population.
   There is a significant selection effect in favor of face-on systems (inclination $\sim0^\circ$) because GW emission is strongest along the direction of the orbital angular momentum vector of the binary.  Modulations in the waveform from spin-induced precession are less detectable for face-on systems.  This selection effect will suppress the ability for precession to break the degeneracy between spin magnitude and mass ratio.}}
   \label{fig:inclination}
\end{figure}

{\bf Because GW emission is strongest along the orbital angular momentum direction, the detected binaries from a population uniform in orientation and volume is biased against systems whose orbital plane is edge-on to the observer (defined as having an inclination angle near $90^\circ$).  Fig.~\ref{fig:inclination} shows the cumulative distribution function of our observed population (red) compared to an underlying population distributed uniformly in orientation. This selection effect has an important role in mass measurement for LV observations.   Precession-induced effects on the waveforms, which are relied upon to break the degeneracy between mass ratio and spin, are less detectable for face-on systems \citep{Vitale:2014mka}.  Previous studies which have used hand-selected populations of signals to study the effects of precession have not accounted for this selection bias, overemphasizing the role precession can play for generic signals.  For example, the inclination used by \citet{Chatziioannou:2014coa} (63$^\circ$) is larger than $\sim80\%$ of our sources so the improvement they found will be fully realized for a small fraction of our simulated population.}

For each binary we compute the response of Advanced LV at design sensitivity \citet{ObservingScenarios} with a low frequency cutoff at 20 Hz.
Each simulated ``detection'' is then analyzed with {\texttt lalinference\_mcmc} which returns independent samples from the PDF for the model parameters.  
We do not simulate instrument noise in this study because our focus is on the degree to which LV's frequency-dependent sensitivity, and the flexibility of the PN waveforms, limit mass measurements.
Adding simulated noise to our signals introduces uncontrollable contributions to the PDF without adding any value to our assessment of each simulated signal.  
Posterior distributions in true detections will be altered by the particular noise realization in which the signal is embedded.  

\section{\label{sec:results} Results}

\subsection{\label{sec:m1m2}Distinguishing black holes from neutron stars}
\begin{figure}
   \centering
   \includegraphics[width=\linewidth]{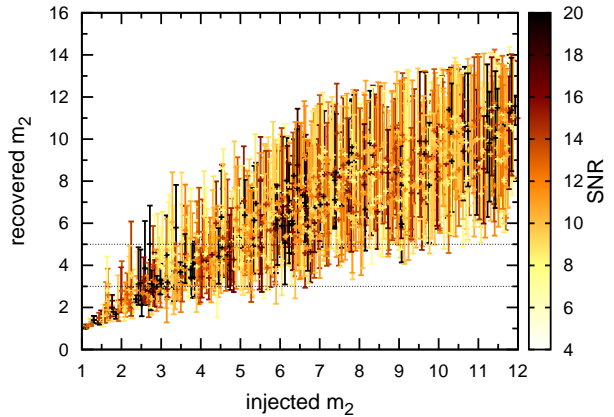} 
   \caption{\small{90\% credible intervals for recovered $m_2$ as a function of the true mass.  Each entry is colored by the network SNR of the source.  Horizontal lines denote the mass gap~\citep{Ozel:2010su, Farr:2010tu}.  Credible intervals for high mass systems are limited from above by our prior on the total mass $M<30$ and the $m_2\leq m_1$ convention.  Axes are in units of $\Msun$.}}
   \label{fig:m2pe}
\end{figure}

We take $m=3\ \Msun$ to be the dividing line between the masses of NSs and BHs, with all NSs located below this threshold and all BHs located above.
We will use $5\ \Msun$ as the minimum mass of a BH when assuming the existence of a mass gap.
The distinction between NSBHs and BBHs in our sample is determined by the measurement of the smaller mass, $m_2$.  
Figure~\ref{fig:m2pe} shows the $90\%$ credible intervals of the PDF for $m_2$ as a function of the true value from the simulated population.
Each entry is colored by the SNR of the signal in the 3-detector network.
Horizontal gray dashed lines denote the mass gap.
We find all of our simulations with $m_2\lesssim1.5\ \Msun$ to be clearly identified as containing at least one neutron star, however our population yields only ten such systems so these are small-number statistics.  Most binaries with a NS below 2 $\Msun$ constrain the NS to have a mass below 3 $\Msun$ although occasional the posterior supports the smaller object being a low-mass BH .
Detections of larger mass neutron stars ($2\leq m_2 \leq 3\ \Msun)$ have $90\%$ credible intervals which consistently extend into the BH regime.
The tendency for recovery of high-mass neutron stars in NSBH systems to be consistent with low-mass BHs poses a challenge for determining the maximum NS mass from LV observations alone.

At what point can we \emph{rule out} the possibility that the system contains a neutron star, and definitively declare that we have detected a binary black hole?
In Figure~\ref{fig:m2pe} we see that the true mass of the smaller object must exceed $\sim 6\ \Msun$ before the $90\%$ credible intervals rule out a NS. {\bf Depending on details of spin alignment and orientation, systems with $m_2$ as low as $4\ \Msun$ can be unambiguously identified as BBHs.}  {\bf Within this range} the $m_2$ posteriors for BBH sources seldom reach below $2\ \Msun$, so if a maximum NS mass were independently confirmed to be consistent with current observations LV's classification of NSs and BHs would improve.

\begin{figure}
   \centering
   \includegraphics[width=\linewidth]{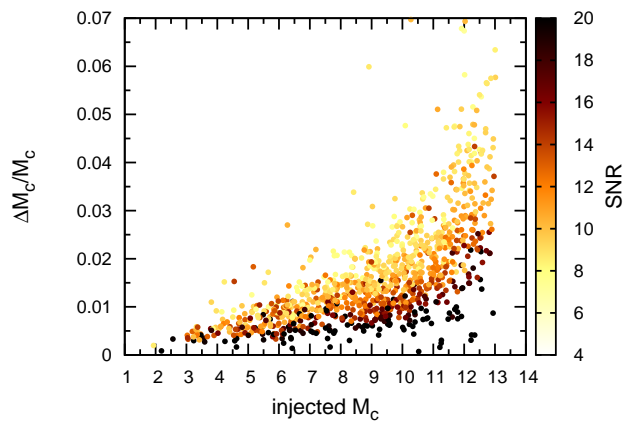} 
   \caption{\small{Fractional $1-\sigma$ error in chirp mass as a function of the true value.  Each entry is colored by the network SNR of the source.  The expected SNR scaling of the errors is evidence in the chirp mass recovery, while it is absent in the $m_2$ measurement due to the degeneracy between mass ratio and spin. The x-axis is in units of $\Msun$.}}
   \label{fig:mcpe}
\end{figure}

{\bf It may seem surprising in Fig~\ref{fig:m2pe} that the width of the credible intervals do not exhibit the $1/{\rm SNR}$ scaling predicted by Fisher matrix approximations~\citep{Cutler:1994ys}.  The Fisher approximation is only suitable at sufficiently high SNR that the posterior distribution function is well approximated by a multivariate Gaussian, in which case the inverse Fisher matrix is the covariance of the posterior.  Implicit in this condition is the assumption that the model waveform is a linear function of the source parameters.  See~\citet{Vallisneri:2007ev} for a thorough deconstruction of Fisher matrix-based intuition being applied to GW signals.  For typical LIGO/Virgo binaries these conditions are not satisfied for the mass parameters.  The width of the $m_2$ credible intervals is driven by uncertainty in $q$ which, due to the degeneracy with spin, is extremely non-Gaussian and can span the entire prior range.  The chirp mass, on the other hand, \emph{is} a sufficiently well constrained parameter at the SNRs in our simulated population.  In Fig.~\ref{fig:mcpe} we show the fractional $1-\sigma$ uncertainty in $\Mc$ inferred from the Markov chains as a function of the source value, with each event colored by the network SNR.  The expected $1/\rm{SNR}$ dependence is apparent for this parameter.  Notice also that the chirp mass errors grow with increasing total mass.  Higher mass binaries are in band for fewer GW cycles which directly impacts measurability as discussed in Sec.~\ref{sec:pe}. }

\subsection{\label{sec:gap}Identifying systems in the mass gap}
Following \cite{Mandel:2010} we use our simulated LV detections to infer the relative fraction of NSs, BHs in the mass gap, and BHs above the mass gap.  
Because of the large mass-measurement uncertainties the number of detections needed to conclude BHs inhabit the mass gap is highly dependent on the underlying mass distribution. Depending on whether the observed population is heavily dominated by low- or high-mass systems, we find 
between ten and many hundreds of detections are necessary to conclude (at three-sigma confidence) the gap is populated.

\begin{figure}
	\centering
   	\includegraphics[width=\linewidth]{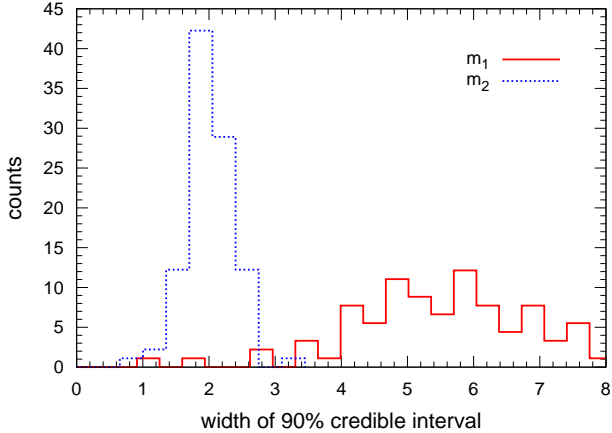} 
  	\caption{\small{Distribution of $90\%$ credible interval widths in $\Msun$ from 100 mass gap sources.  The width of the mass gap is 2 $\Msun$, and therefore most of the BHs cannot be constrained to fall within the gap.}}
	\label{fig:m2_gap}
\end{figure}

Our initial set of simulated signals features many high-mass ratio binaries.  We simulated an additional 100 sources with both objects having $3<m<5\ \Msun$ to check whether comparable-mass systems are easier to identify in the gap. The mass-gap population still suffers from large mass errors with $>95\%$ of the sources having posterior support for a NSBH system.
Figure~\ref{fig:m2_gap} shows the distribution of the 90\% credible interval widths for $m_1$ (red, solid) and $m_2$ (blue, dotted) of the gap sources.  
The majority of plausible mass-gap sources yield credible intervals that are similar to or exceed the width of the mass gap.

However, $\sim 25\%$ of the mass gap sources' $m_2$ posteriors do not reach below $2\ \Msun$.
While we will not be able to say with any certainty that an individual source occupies the mass gap, we can often conclude that the binary contains 
either an unusually high-mass neutron star, or a pair of unusually low-mass black holes.

Three of the mass-gap sources were constrained to be $3<m<5\ \Msun$.  A careful investigation of these systems revealed they were in low-probability alignments -- all with the spin of the larger mass close to the orbital plane, and the best constrained having $\vec L$ nearly perpendicular to the line of sight. 

\subsection{\label{sec:spin}Restricting allowed spins for neutron stars}

All of the results thus far presented used a uniform prior on spin magnitude between $[0,1]$.
The maximum upper limit on the spin of a NS is $\chi\lesssim0.7$ while observed NS spins are lower~\citep{Lo:2011}.
We investigate if using a physical prior on NS spins can improve $m_2$ measurement.

We resample the posterior by imposing a maximum spin for component masses below $3\ \Msun$, rejecting samples from the Markov chain with $\chi_2>0.7$.  
We found no indication that restricting the range of neutron star spins significantly improves uncertainty in the determination of $m_2$.

\begin{figure}
   \centering
   \includegraphics[width=\linewidth]{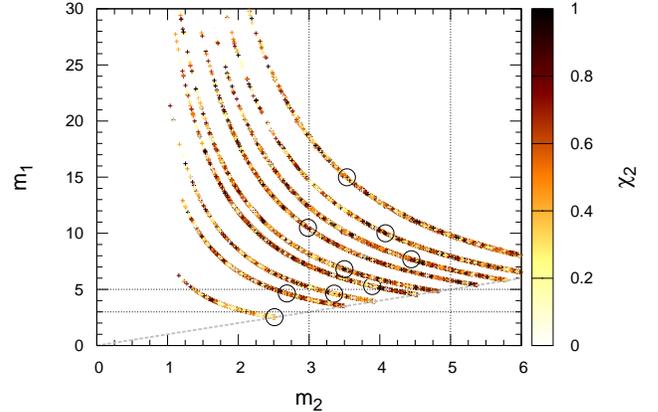} 
   \caption{\small{Component mass recovery colored by the spin of the lower mass.  Open circles mark the true values for nine representative binary systems from the simulated population.  The spin of the lower mass is not well constrained, and does not strongly correlate with the mass parameters.  Therefore, restricting the spin of the smaller body to be consistent with our prior expectations does not impact our ability to distinguish between black holes and neutron stars. Both x- and y-axes are in units of $\Msun$.}}
   \label{fig:m1m2a2}
\end{figure}

Figure~\ref{fig:m1m2a2} demonstrates the limited role of $m_2$'s spin on mass determination.  
Open circles mark the true values of component masses for nine representative examples.  
Going through each circle is the scatter plot of the posterior samples colored by  $\chi_2$. 
The arcs traced out by the samples are lines of constant $\Mc$.
The vertical and horizontal dashed lines denote the mass gap.
Notice that there is no obvious correlation between position along the arc and $\chi_2$ -- the spin of the smaller body is generally not constrained and therefore does not help with mass determination.  
Restricting the spin of the smaller mass does not help because at high mass ratios (and therefore NS-like $m_2$), the contribution to the PN phase from $\chi_2$ is suppressed.  
The leading order spin corrections enter the PN phase with magnitude $\chi_i m_i^2$, so $\chi_2$'s influence to the phase evolution is down-weighted relative to $\chi_1$ by $\mathcal{O}\left(({\rm mass\ ratio})^2\right)$.

\section{\label{sec:discuss} Discussion}

In this paper we investigated the capability of the Advanced LV network to distinguish between NSs and BHs {\bf from the inspiral-only waveforms} using a large population of plausible detections.  
This is the first large-scale study to characterize compact binary PDFs including spin-precession effects over a broad range of masses, mass ratios, and spins.  Our study does not factor in systematic effects from real detector noise or differences between template waveforms and the true gravitational wave signal.
{\bf Our study is limited to inspiral-only waveforms for simulation of signals and template waveforms for recovery because available precessing merger/ringdown models are not valid over the full prior volume.  For many of our simulated signals the merger will be detectable and may help improve component mass estimates.  Further improvements may come from including PN amplitude corrections.}  

We arrive at four main conclusions from our analysis:
\begin{enumerate}[leftmargin=*]
\item \emph{When are we certain of at least one NS?}  For most systems with $m_2 \leq 2\ \Msun$, and all systems with $m_2 \leq1 .5\ \Msun$.  
  \end{enumerate}
For larger-mass neutron stars ($2 \leq m_2 \leq 3$) the 90\% credible intervals frequently extend into the low-mass BH regime. This tendency for high-mass neutron stars in NSBH systems to be consistent with low-mass black holes poses a challenge for determining the maximum NS mass from LV observations alone.
  \begin{enumerate}[resume,leftmargin=*]  
  \item \emph{When are we certain of a BBH with both masses above 3 $\Msun$?} When the mass of the smaller object exceeds $\sim 6\ \Msun$.
  \end{enumerate}
  The $m_2$ posteriors for BBH signals seldom reach below 2 $\Msun$, so if a maximum neutron star mass were independently confirmed to be $m_{\rm max}\sim2\Msun$, then LV's ability to discriminate between BBH and BNS/NSBH populations would be significantly improved.
  \begin{enumerate}[resume,leftmargin=*]  
  \item \emph{When is a black hole definitely not in the mass gap?} When component mass $m_2 \gtrsim 10\ \Msun$ the 90\% credible intervals do not reach into the mass gap.
  \end{enumerate}
 It may prove challenging to confirm its existence because NSs with masses above $\sim 2\ \Msun$ have error bars reaching into the $m_2\in[3,5]\ \Msun$ interval from below while BBH systems with $m_2 \lesssim 10\ \Msun$ can have credible intervals that extend below $5\ \Msun$.  Our findings suggest that inferring a bimodal mass distribution based solely on the observed sample of compact binaries especially without reliable predictions for the underlying compact object mass distribution will be challenging.  
  \begin{enumerate}[resume,leftmargin=*]  
  \item \emph{When are we certain of a mass-gap object?} Only in rare circumstances when the binary is nearly edge-on and spins are oriented in the orbital plane.  For typical binaries component mass errors are larger than the mass gap.  
  \end{enumerate}
We simulated an additional 100 injections with both masses in the mass gap between $[3,5]$ $\Msun$.  We found $> 95\%$ of the systems had some posterior support for a NSBH system.  The majority of plausible mass gap sources yield credible intervals that are similar to, or exceed the width of the mass gap itself (2 $\Msun$).  
{\bf Only sources in low-probability alignments (spins in the orbital plane and/or edge-on orientations) were constrained to be $3<m<5\ \Msun$.}
However, $m_2$ posteriors from mass gap BBHs do not reach below 2 $\Msun$. While we will not be able to say with any certainty that a given source occupies the mass gap, we will at least be alerted to the fact that the binary contains either an unusually high-mass neutron star, or a pair of unusually low mass black holes, with either possibility providing plenty of intrigue for the astrophysics community.

Assuming flat mass distributions, we used simulated LV detections to infer the relative fraction of NSs and BHs in and above the mass gap, and estimate how many detections would be needed to confidently conclude that gap BHs exist. Depending on whether the underlying mass distribution is heavily dominated by low- or high-mass systems, ten to hundreds of detections are needed to confirm objects in the gap. Assuming a plausible population-synthesis model where BNSs, NSBHs, and BBHs are well separated in chirp-mass space, \citet{Mandel:2015spa} found that a mass gap was statistically distinguishable with a only few tens of detections
It is clear that forecasts for what we may learn from LV observations of compact binaries are subject to large variance while uncertainty about the true mass distribution persists.
Continued effort in theoretically understanding that distribution, and how LIGO/Virgo observations can be used to test such theories, will be of great value as we begin assembling a catalog of compact binaries coalescences.

\section{Acknowledgments}
We thank Atul Adhikari, Claudeson Azurin, Brian Klein, Brandon Miller, Leah Perri, Ben Sandeen, Jeremy Vollen, Michael Zevin for help running the MCMC analysis.  Will Farr, Carl-Johan Haster, and Ilya Mandel provided important comments about our calculations and results.
TBL and VK acknowledge NSF award PHY-1307020.  BF was supported by the Enrico Fermi Institute at the University of Chicago as a McCormick Fellow. SC thanks the US-UK Fulbright Commission for personal financial support during this research period.  DEH acknowledges NSF CAREER grant PHY-1151836, the Kavli Institute for Cosmological Physics at the University of Chicago through NSF grant PHY-1125897, and an endowment from the Kavli Foundation.  Computational resources were provided by the Northwestern University Grail cluster through NSF MRI award PHY-1126812.

%

\end{document}